\documentclass[aps,prd,twocolumn,superscriptaddress,preprintnumbers]{revtex4-1}
\usepackage{graphicx}
\usepackage{amsmath}
\usepackage{amssymb}
\usepackage{bm}
\usepackage{epsfig}
\begin{document}

\title{\boldmath 
Production and polarization of $S$-wave quarkonia in 
potential nonrelativistic QCD
}
\preprint{TUM-EFT 168/22}

\author{Nora~Brambilla}
\affiliation{Physik Department, Technische Universit\"at M\"unchen,\\
James-Franck-Strasse 1, 85748 Garching, Germany} 
\affiliation{Institute for Advanced Study, Technische Universit\"at
M\"unchen,\\ Lichtenbergstrasse 2 a, 85748 Garching, Germany} 
\affiliation{Munich Data Science Institute,
Technische Universit\"at M\"unchen, \\ Walther-von-Dyck-Strasse 10, 85748
Garching, Germany}
\author{Hee~Sok~Chung}
\altaffiliation[Present address:]{~Department of Physics, Korea University, 
Seoul 02841, Korea}
\affiliation{Physik Department, Technische Universit\"at M\"unchen,\\
James-Franck-Strasse 1, 85748 Garching, Germany} 
\affiliation{Excellence Cluster ORIGINS, Boltzmannstrasse
2, 85748 Garching, Germany}
\author{Antonio~Vairo}
\affiliation{Physik Department, Technische Universit\"at M\"unchen,\\
James-Franck-Strasse 1, 85748 Garching, Germany} 
\author{Xiang-Peng~Wang}
\affiliation{Physik Department, Technische Universit\"at M\"unchen,\\
James-Franck-Strasse 1, 85748 Garching, Germany} 
\noaffiliation
\date{\today}

\begin{abstract}
Based on the potential nonrelativistic QCD formalism, we
compute the nonrelativistic QCD long-distance matrix elements (LDMEs)
for inclusive production of $S$-wave heavy quarkonia. 
This greatly reduces the number of nonperturbative unknowns and
brings in a substantial enhancement in the predictive power of the 
nonrelativistic QCD 
factorization formalism. We obtain improved determinations of the LDMEs 
and find cross sections and polarizations of $J/\psi$, $\psi(2S)$, and excited
$\Upsilon$ states that agree well with LHC data. 
Our results may have important implications in pinning down the
heavy quarkonium production mechanism.
\end{abstract}

\maketitle

Production of heavy quarkonia in high energy collisions provide a unique
opportunity to probe the interplay between perturbative and
nonperturbative aspects of QCD, as well as the hot and dense phase of
QCD~\cite{Brambilla:2004wf,Brambilla:2010cs,Bodwin:2013nua,Brambilla:2014jmp}.
Especially, inclusive production rates of $S$-wave heavy quarkonia 
such as $J/\psi$, $\psi(2S)$, and $\Upsilon$ have been
studied extensively in collider experiments such as the RHIC, Tevatron, $B$
factories, HERA, and the LHC, and will continue to be an important subject in future
colliders including the Electron-Ion Collider. 

Phenomenological studies of heavy quarkonium production 
have mostly been carried out in the context of nonrelativistic effective 
field theories, which make use of the hierarchy of energy scales $m \gg mv
\gg mv^2$ associated with a heavy quarkonium state. Here, $m$ is the heavy
quark mass, and $v$ is the velocity of a heavy quark in the heavy quarkonium
rest frame. 
Nonrelativistic QCD (NRQCD) follows from integrating out the scale of
the heavy quark mass $m$~\cite{Caswell:1985ui,Bodwin:1994jh}. 
NRQCD provides a factorization formalism that
describes the inclusive cross section of a heavy quarkonium in terms
of sums of products of perturbatively calculable short-distance coefficients
(SDCs) and nonperturbative long-distance matrix elements (LDMEs). 
The LDMEs have known scalings in $v$, so that in practice the factorization 
formula is truncated at a desired accuracy in $v$. NRQCD factorization 
for inclusive production is expected to be accurate up to relative order 
$m^2/p_T^2$ in the expansion in powers of $m^2/p_T^2$, where $p_T$ is 
the transverse momentum of the quarkonium produced in the
collision~\cite{Bodwin:1994jh, Nayak:2005rw, Nayak:2005rt, Nayak:2006fm, 
Kang:2014tta}. 
Hence, large-$p_T$ cross sections of heavy quarkonia are described
by the NRQCD factorization formalism in terms of a limited number of LDMEs, 
which depend only on the nonperturbative nature of the heavy quarkonium state 
but are process independent. 

For decades a huge effort has been put into computing the SDCs and
determining the LDMEs. 
As it has not been known how to compute the LDMEs from first principles, 
the determinations of the LDMEs have
mostly relied on measured cross section data. This approach has led to
inconsistent sets of LDMEs that give contradicting predictions, 
depending on the choice of data and the organization of the 
QCD perturbation series~\cite{Ma:2010jj, Butenschoen:2010rq, Ma:2010yw,
Chao:2012iv, Butenschoen:2012px, Gong:2012ug, Butenschoen:2011yh,
Butenschoen:2009zy, Butenschoen:2011ks, Bodwin:2014gia, Bodwin:2015iua}. 
Also the signs of the LDMEs can differ between different determinations. 
As none of the existing 
determinations are able to give a comprehensive
description of important observables, it is fair to say that the production
mechanism of heavy quarkonium still remains elusive~\cite{Chung:2018lyq}. 

Recently, 
a formalism for computing the production LDMEs has been developed
in Refs.~\cite{Brambilla:2020ojz, Brambilla:2021abf}
based on the effective field theory potential NRQCD (pNRQCD),
which is obtained by integrating out the scale of order
$mv$~\cite{Pineda:1997bj,Brambilla:1999xf,Brambilla:2004jw}. 
The pNRQCD formalism provides expressions for the LDMEs in terms of
quarkonium wave functions at the origin, which can be computed by solving the
Schr\"odinger equation, and gluonic correlators, which are universal quantities
that do not depend on the specific heavy quarkonium state 
and can in principle be computed in lattice QCD. 
This results in a reduction of the number of nonperturbative unknowns 
which greatly enhances the predictive power of NRQCD factorization. 
The pNRQCD formalism has been
successfully applied to the production of $P$-wave heavy quarkonia, and the
phenomenological results agree well with available measurements at the
LHC~\cite{Brambilla:2020ojz, Brambilla:2021abf}. 
It has been anticipated that the application of the pNRQCD formalism to 
production of $S$-wave heavy quarkonia may help scrutinize the
LDMEs and the applicability of the NRQCD factorization formalism. 

In this work we compute, for the first time, the NRQCD LDMEs for production
of $S$-wave heavy quarkonia in the pNRQCD formalism. We work in the strongly
coupled regime, $v \gtrsim \Lambda_{\rm QCD}/m$, which we assume to be
appropriate for $J/\psi$, $\psi(2S)$, and excited
$\Upsilon$ states. 
Based on the results for the LDMEs that we obtain, 
we compute production rates of $S$-wave quarkonia at the LHC 
and compare them with data.

The inclusive cross section of a spin-1 $S$-wave heavy quarkonium $V$ is 
given in the NRQCD factorization formalism at relative order $v^4$ accuracy by
\begin{align}
\label{eq:NRQCDfac}
\sigma_{V+X} &= 
\hat{\sigma}_{{}^3S_1^{[1]}} 
\langle {\cal O}^{V} ({}^3S_1^{[1]}) \rangle 
+ \hat{\sigma}_{{}^3S_1^{[8]}} 
\langle {\cal O}^{V} ({}^3S_1^{[8]}) \rangle 
\nonumber \\ & \quad 
+ \hat{\sigma}_{{}^1S_0^{[8]}} 
\langle {\cal O}^{V} ({}^1S_0^{[8]}) \rangle 
+ \hat{\sigma}_{{}^3P_J^{[8]}} 
\langle {\cal O}^{V} ({}^3P_0^{[8]}) \rangle. 
\end{align}
Here, the $\hat{\sigma}_{N}$ are the SDCs, which correspond to the 
production rate of a heavy quark $Q$ and antiquark $\bar Q$ 
in the color and angular momentum state $N$. 
The SDC $\hat{\sigma}_{{}^3P_J^{[8]}}$ includes contributions from 
$J=0$, 1, and 2.
We use spectroscopic notation for the angular momentum state of the $Q
\bar Q$, while the superscripts $[1]$ and $[8]$ denote the color state of the 
$Q \bar Q$: color singlet (CS) and color octet (CO), respectively.
The LDMEs are defined by~\cite{Bodwin:1994jh, Nayak:2005rw, Nayak:2005rt, 
Nayak:2006fm}
\begin{subequations}
\label{eq:matrix_elements}
\begin{align}
\langle {\cal O}^{V} ({}^3S_1^{[1]}) \rangle &= \langle \Omega |
\chi^\dag \sigma^i \psi {\cal P}_{V(\bm{P}=\bm{0})}
\psi^\dag \sigma^i \chi | \Omega \rangle,
\\
\langle {\cal O}^{V} ({}^3S_1^{[8]}) \rangle &= \langle \Omega |
\chi^\dag \sigma^i T^a \psi \Phi_\ell^{\dag ab} {\cal
P}_{V(\bm{P}=\bm{0})}
\nonumber \\ & \quad \times \Phi_\ell^{bc} \psi^\dag \sigma^i T^c \chi | \Omega \rangle,
\\
\langle {\cal O}^{V} ({}^1S_0^{[8]}) \rangle
&= \langle \Omega |
\chi^\dag T^a \psi \Phi_\ell^{\dag ab} {\cal
P}_{V(\bm{P}=\bm{0})}
\nonumber \\ & \quad \times \Phi_\ell^{bc} \psi^\dag T^c \chi | \Omega \rangle,
\\
\langle {\cal O}^{V} ({}^3P_0^{[8]}) \rangle &= \frac{1}{3} \langle \Omega | 
\chi^\dag (- \tfrac{i}{2} \overleftrightarrow{\bm{D}} \cdot
\bm{\sigma}) T^a \psi \Phi_\ell^{\dag ab} {\cal P}_{V(\bm{P}=\bm{0})}
\nonumber \\ & \quad \times \Phi_\ell^{bc} \psi^\dag (- \tfrac{i}{2}
\overleftrightarrow{\bm{D}} \cdot \bm{\sigma}) T^c \chi | \Omega \rangle,
\end{align}
\end{subequations}
where $| \Omega \rangle$ is the QCD vacuum, $T^a$ are SU(3) generators,
$\sigma^i$ are Pauli matrices, 
and $\psi$ and $\chi$ are Pauli spinors that annihilate and create a heavy quark and antiquark, respectively. 
The covariant derivative $\overleftrightarrow{\bm{D}}$ is defined by 
$\chi^\dag \overleftrightarrow{\bm{D}} \psi = \chi^\dag \bm{D} \psi 
- (\bm{D} \chi)^\dag \psi$, 
with $\bm{D} = \bm{\nabla} -i g \bm{A}$, and $\bm{A}$ is the gluon field. 
The operator ${\cal P}_{{\cal Q} (\bm{P})}$ projects onto a state consisting of a quarkonium $\cal Q$ with momentum $\bm{P}$.
The path-ordered Wilson line along the spacetime direction $\ell$, defined by $\Phi_\ell = {\cal P} \exp [ -i g \int_0^\infty d\lambda\, \ell \cdot A^{\rm adj}(\ell \lambda) ]$,
where $A^{\rm adj}$ is the gluon field in the adjoint representation, ensures
the gauge invariance of the CO LDMEs~\cite{Nayak:2005rw, Nayak:2005rt,
Nayak:2006fm}. 
The direction $\ell$ is arbitrary. 

In existing studies of $S$-wave heavy quarkonium production, the CO LDMEs are
determined by comparing Eq.~(\ref{eq:NRQCDfac}) to cross section data. 
One major difficulty in this approach is that 
in existing studies based on hadroproduction,
only certain linear combinations of the CO LDMEs are strongly
constrained, while individual LDMEs are often poorly determined. 

Now we compute the LDMEs in Eqs.~(\ref{eq:matrix_elements}) in pNRQCD using the
formalism developed in Refs.~\cite{Brambilla:2020ojz, Brambilla:2021abf}. 
We work at leading nonvanishing order in the 
quantum-mechanical perturbation theory (QMPT), where we expand in powers of
$v^2$ and $\Lambda_{\rm QCD}/m$. 

For the CS LDME $\langle {\cal O}^{V} ({}^3S_1^{[1]}) \rangle$, 
we obtain at leading order in QMPT
\begin{equation}
\label{eq:singlet_result}
\langle {\cal O}^{V} ({}^3S_1^{[1]}) \rangle 
= 2 N_c \times \frac{3 | R^{(0)}_V (0)|^2}{4 \pi},
\end{equation}
where $R_{V}^{(0)}(r)$ is the radial wave function of the quarkonium $V$ at
leading order in $v$.
This reproduces the result obtained in the vacuum-saturation approximation in Ref.~\cite{Bodwin:1994jh}. 

The expressions for the CO LDMEs are given by 
\begin{subequations}   
\label{eq:octet_result}
\begin{align}   
\langle {\cal O}^{V} ({}^3S_1^{[8]}) \rangle &= 
\frac{1}{2 N_c m^2} \frac{3 | R^{(0)}_V (0)|^2}{4 \pi} {\cal E}_{10;10}, 
\\
\langle {\cal O}^{V} ({}^1S_0^{[8]}) \rangle &= 
\frac{1}{6 N_c m^2} \frac{3 | R^{(0)}_V (0)|^2}{4 \pi} c_F^2 {\cal B}_{00}, 
\\
\langle {\cal O}^{V} ({}^3P_0^{[8]}) \rangle &= 
\frac{1}{18 N_c} \frac{3 | R^{(0)}_V (0)|^2}{4 \pi} {\cal E}_{00}, 
\end{align}
\end{subequations}   
where $c_F$ is given in Refs.~\cite{Eichten:1990vp, Czarnecki:1997dz, 
Grozin:2007fh} in the $\overline{\rm MS}$ scheme at the scale
$\Lambda$ by $c_F = 1+ \frac{\alpha_s}{2 \pi} [ C_F + C_A (1 + \log \Lambda/m) ] 
+ O(\alpha_s^2)$, 
with $C_F = (N_c^2-1)/(2 N_c)$ and $C_A = N_c$; 
${\cal E}_{10;10}$,
${\cal B}_{00}$, and ${\cal E}_{00}$ are gluonic correlators of dimension 2
defined by 
\begin{subequations}
\label{eq:correlators}
\begin{align}
{\cal E}_{10;10} &= 
\Big|
d^{d a c}
\int_0^\infty dt_1 \, t_1 \int_{t_1}^\infty dt_2 \,
g E^{b,i} (t_2) 
\nonumber \\ & \hspace{2ex} \times 
\Phi_0^{b c} (t_1;t_2) 
g E^{a,i} (t_1) \Phi_0^{d f} (0;t_1) \Phi_\ell^{ef}
|\Omega\rangle \Big|^2,
\\
{\cal B}_{00} &= 
\Big| \int_0^\infty dt  \, 
g {B}^{a,i} (t) \Phi_0^{ac} (0;t) \Phi_\ell^{bc} | \Omega \rangle \Big|^2,
\\
{\cal E}_{00} &= 
\Big| \int_0^\infty dt \, 
g {E}^{a,i} (t) \Phi_0^{ac} (0;t) \Phi_\ell^{bc} | \Omega \rangle \Big|^2,
\end{align}
\end{subequations}
where $d^{abc} = 2 \, {\rm tr} (\{ T^a, T^b \} T^c)$, 
$E^{a,i}(t)$ and $B^{a,i}(t)$
are chromoelectric and chromomagnetic field components, respectively, 
computed at the time $t$ and space coordinate $\bm{0}$, 
and $\Phi_0 (t,t') = {\cal P} \exp [ -i g \int_t^{t'} d\tau\, A_0^{\rm
adj}(\tau,\bm{0}) ]$ is a Schwinger line.
The chromoelectric and chromomagnetic fields in the expressions for 
${\cal E}_{10;10}$ and ${\cal B}_{00}$ come from the 
$\bm{D}^2$ and $\bm{\sigma} \cdot g \bm{B}$ terms in the NRQCD Lagrangian, 
respectively, 
while the chromoelectric fields in ${\cal E}_{00}$ come from the
$\overleftrightarrow{\bm{D}}$ in the definition of $\langle {\cal O}^{V}
({}^3P_0^{[8]}) \rangle$.
Note that the correlators ${\cal E}_{10;10}$, ${\cal B}_{00}$, and 
${\cal E}_{00}$ 
are purely gluonic quantities that do not depend on the heavy quark flavor. 
The expressions in Eqs.~(\ref{eq:octet_result}) are accurate up to corrections
of relative order $v^2$ and $1/N_c^2$~\cite{Brambilla:2020ojz,
Brambilla:2021abf}. 

We find the one-loop evolution equation for ${\cal E}_{10;10}$ given by 
\begin{equation}
\label{eq:RG_E}
\frac{d}{d \log \Lambda} {\cal E}_{10;10} = 
{\cal E}_{00} \times \frac{2 \alpha_s}{3 \pi} \frac{N_c^2-4}{N_c} +
O(\alpha_s^2). 
\end{equation}
This reproduces the 
known scale dependence of $\langle {\cal O}^{V} ({}^3S_1^{[8]}) \rangle$, 
which cancels the explicit 
$\log \Lambda$ that appears in 
$\hat{\sigma}_{^3P_J^{[8]}}$~\cite{Bodwin:1994jh, Bodwin:2012xc}.
The correlator ${\cal B}_{00}$ also depends on the scale in a way that 
$c_F^2 {\cal B}_{00}$ is scale independent at one-loop level. 

\begin{table}
\begin{ruledtabular}
\begin{tabular}{c|ccc}
$p_T$ region & ${\cal E}_{10;10}$ (GeV$^2$) 
& $c_F^2 {\cal B}_{00}$ (GeV$^2$)
& ${\cal E}_{00}$ (GeV$^2$) 
\\
\hline
$p_T/(2 m) > 5$ &   0.860 $\pm$ 0.277  & $-$2.25 $\pm$ 7.06 & 13.4 $\pm$ 4.6 
\\
$p_T/(2 m) > 3$ &   1.17 $\pm$ 0.13 & $-$9.79 $\pm$ 3.08 & 18.5 $\pm$ 2.1 
\end{tabular}
\caption{Fit results for the correlators 
${\cal E}_{10;10}$, $c_F^2 {\cal B}_{00}$, and ${\cal E}_{00}$
for the 
two $p_T$ regions in the $\overline{\rm MS}$ scheme at the scale $\Lambda =
1.5$~GeV. The SDC $c_F$ is computed for the charm quark mass $m=1.5$~GeV. 
}
\label{tab:1}
\end{ruledtabular}
\end{table}

\begin{table}
\begin{ruledtabular}
\begin{tabular}{c|ccc}
$p_T$ region 
& $\langle {\cal O}^{J/\psi} (^3S_1^{[8]}) \rangle$ 
& $\langle {\cal O}^{J/\psi} (^1S_0^{[8]}) \rangle$ 
& $\langle {\cal O}^{J/\psi} (^3P_0^{[8]}) \rangle/m^2$ 
\\
\hline
$p_T/(2 m) > 5$ &  1.25 $\pm$ 0.40  & $-$1.10 $\pm$ 3.43 & 2.18 $\pm$ 0.75
\\
$p_T/(2 m) > 3$ &  1.70 $\pm$ 0.18 & $-$4.76 $\pm$ 1.50 & 3.00 $\pm$ 0.34
\end{tabular}
\caption{
Numerical results for the $J/\psi$ CO LDMEs in units of $10^{-2}$~GeV$^3$. 
The uncertainties come from the correlators ${\cal E}_{10;10}$, $c_F^2 {\cal
B}_{00}$, and ${\cal E}_{00}$. 
}
\label{tab:2}
\end{ruledtabular}
\end{table}

We note that our results in Eqs.~\eqref{eq:octet_result} are valid in
dimensional regularization (DR), because in computing the LDMEs we have 
discarded scaleless power divergences which vanish in
DR~\cite{Brambilla:2002nu, Brambilla:2021abf}. 
Since the correlators in Eqs.~(\ref{eq:correlators}) contain power 
UV divergences which are automatically subtracted in DR, 
they may not be positive definite, 
even though they are defined as norms of states that are obtained by applying
gluonic operators on the QCD vacuum.

The three correlators in Eqs.~(\ref{eq:correlators})
and $|R^{(0)}_V(0)|$ completely fix all LDMEs 
that appear in Eq.~(\ref{eq:NRQCDfac}). 
Since lattice determinations of the correlators are not available yet, 
we determine the correlators by comparing 
Eq.~(\ref{eq:NRQCDfac}) to measured cross section data at the LHC.
We employ the measured data 
for prompt $J/\psi$ and $\psi(2S)$ 
production rates in Refs.~\cite{CMS:2011rxs, CMS:2015lbl} 
and 
inclusive $\Upsilon(2S)$ and $\Upsilon(3S)$ cross sections in
Ref.~\cite{ATLAS:2012lmu}. 
We use $p_T$ cuts to prevent factorization-breaking effects at
small $p_T$ from affecting the fit.
In order to explore the dependence on the $p_T$ cut, we consider two 
regions $p_T/(2 m) > 5$ and $p_T/(2 m) > 3$.

In computing the cross sections from Eqs.~(\ref{eq:NRQCDfac}) and 
(\ref{eq:octet_result}), we use the values 
$|R^{(0)}_{J/\psi} (0)|^2 = 0.825$~GeV$^3$, 
$|R^{(0)}_{\psi(2S)} (0)|^2 = 0.492$~GeV$^3$, 
$|R^{(0)}_{\Upsilon(2S)} (0)|^2 = 3.46$~GeV$^3$, 
and 
$|R^{(0)}_{\Upsilon(3S)} (0)|^2 = 2.67$~GeV$^3$, 
which we obtain by comparing the measured leptonic decay rates in 
Ref.~\cite{Ablikim:2012xi} with the pNRQCD expressions at leading order in $v$
and at next-to-leading order (NLO) in $\alpha_s$~\cite{Brambilla:2020xod}. 
We compute the SDCs in Eq.~(\ref{eq:NRQCDfac}) at NLO in $\alpha_s$ 
by using the \textsc{FDCHQHP} package~\cite{Wan:2014vka}.
We use $m = 1.5$~GeV for charm and $4.75$~GeV for bottom, and set 
$\Lambda = m$ for the $\overline{\rm MS}$ scale of the NRQCD LDMEs. 
We consider the runnings of the correlators by using the 
renormalization-group improved formulas at one loop; the effect of
the running of ${\cal B}_{00}$ is numerically small, while the running of 
${\cal E}_{10;10}$ depends on the value of ${\cal E}_{00}$. 
We consider feeddowns from $P$-wave quarkonia by using the measurements 
in Refs.~\cite{ATLAS:2014ala, LHCb:2014ngh}, 
and take into account the decays of $\psi(2S)$ into $J/\psi$ and 
$\Upsilon(3S)$ into $\Upsilon(2S)$
using the measured branching ratios in Ref.~\cite{Tanabashi:2018oca}. 
We take the uncertainties in the theoretical expressions for the charmonium
and bottomonium cross sections to be 30\% and 10\% of the central values,
respectively, which account for uncalculated corrections of higher orders 
in $v^2$. We neglect uncertainties from corrections of order 
$1/N_c^2$ and variations of scales because they are small compared to
the uncertainties that we consider.

The pNRQCD results for the LDMEs imply that the ratio of the
large-$p_T$ direct production rates of $\psi(2S)$ and $J/\psi$ is simply 
given by 
$|R^{(0)}_{\psi(2S)} (0)|^2/|R^{(0)}_{J/\psi} (0)|^2$, up to
corrections of order $v^2$, independently of the values of the correlators. 
If we take into account the feeddown contributions from decays of $\chi_{c}$
and $\psi(2S)$ into $J/\psi$, we obtain 
$(B_{\psi(2S) \to \mu^+ \mu^-} \times \sigma_{\psi(2S)+X})/
(B_{J/\psi \to \mu^+ \mu^-} \times \sigma_{J/\psi+X})
\approx 0.04$, which agrees well with the measured values in
Ref.~\cite{CMS:2011rxs} at large $p_T$. 
Similarly, the ratio of the large-$p_T$ direct production rates of 
$\Upsilon(3S)$ and
$\Upsilon(2S)$ is given by $|R^{(0)}_{\Upsilon(3S)} (0)|^2/
|R^{(0)}_{\Upsilon(2S)} (0)|^2$, up to corrections of order $v^2$. 
If we take into account the feeddowns from
decays of $\chi_{b} (3P)$ into $\Upsilon(3S)$, as well as 
$\chi_{b}(3P)$, 
$\chi_{b} (2P)$, and $\Upsilon(3S)$ into $\Upsilon(2S)$, 
we obtain 
$(B_{\Upsilon(3S) \to \mu^+ \mu^-} \times \sigma_{\Upsilon(3S)+X})/
(B_{\Upsilon(2S) \to \mu^+ \mu^-} \times \sigma_{\Upsilon(2S)+X})
\approx 0.8$, which is in fair agreement with measurements in
Ref.~\cite{ATLAS:2012lmu}. 

The values of the correlators that are determined from our fits are listed in
Table~\ref{tab:1}. The results from the two $p_T$ regions are consistent
within uncertainties. 
The qualities of the fits are good; 
we obtain $\chi^2_{\rm min}/{\rm d.o.f.} = 6.30/41$ and $14.0/71$ for 
$p_T/(2 m) > 5$ and $p_T/(2 m) > 3$, respectively.
As a representative example, we show the numerical results for the 
$J/\psi$ CO LDMEs in Table~\ref{tab:2}.
The uncertainties shown in Table~\ref{tab:1} are correlated; the correlation
matrix of the uncertainties in ${\cal E}_{10;10}$, $c_F^2 {\cal B}_{00}$, and 
${\cal E}_{00}$ 
for each $p_T$ region is given by 
\begin{subequations}
\label{eq:covmat}
\begin{align}
C_{\frac{p_T}{2 m}>5} &= 
\begin{pmatrix}
 0.0766 & -1.75 & \phantom{-}1.27 \\
 -1.75 & \phantom{-}49.8 & -28.5 \\
\phantom{-} 1.27 & -28.5 & \phantom{-}21.3 
\end{pmatrix} \textrm{~GeV}^4, 
\\
C_{\frac{p_T}{2 m}>3} &=
\begin{pmatrix}
 0.0160 & -0.348 & 0.267 \\
 -0.348 & \phantom{-}9.49 & -5.62 \\
\phantom{-} 0.267 & -5.62 & \phantom{-}4.48 
\end{pmatrix}\textrm{~GeV}^4.
\end{align}
\end{subequations}
When computing observables, we take into account these 
correlations in order to reduce the theoretical uncertainties.
We note that the eigenvectors of the correlation matrix are
almost independent of the $p_T$ cut. 

\begin{figure}[t]
\includegraphics[width=0.49\textwidth]{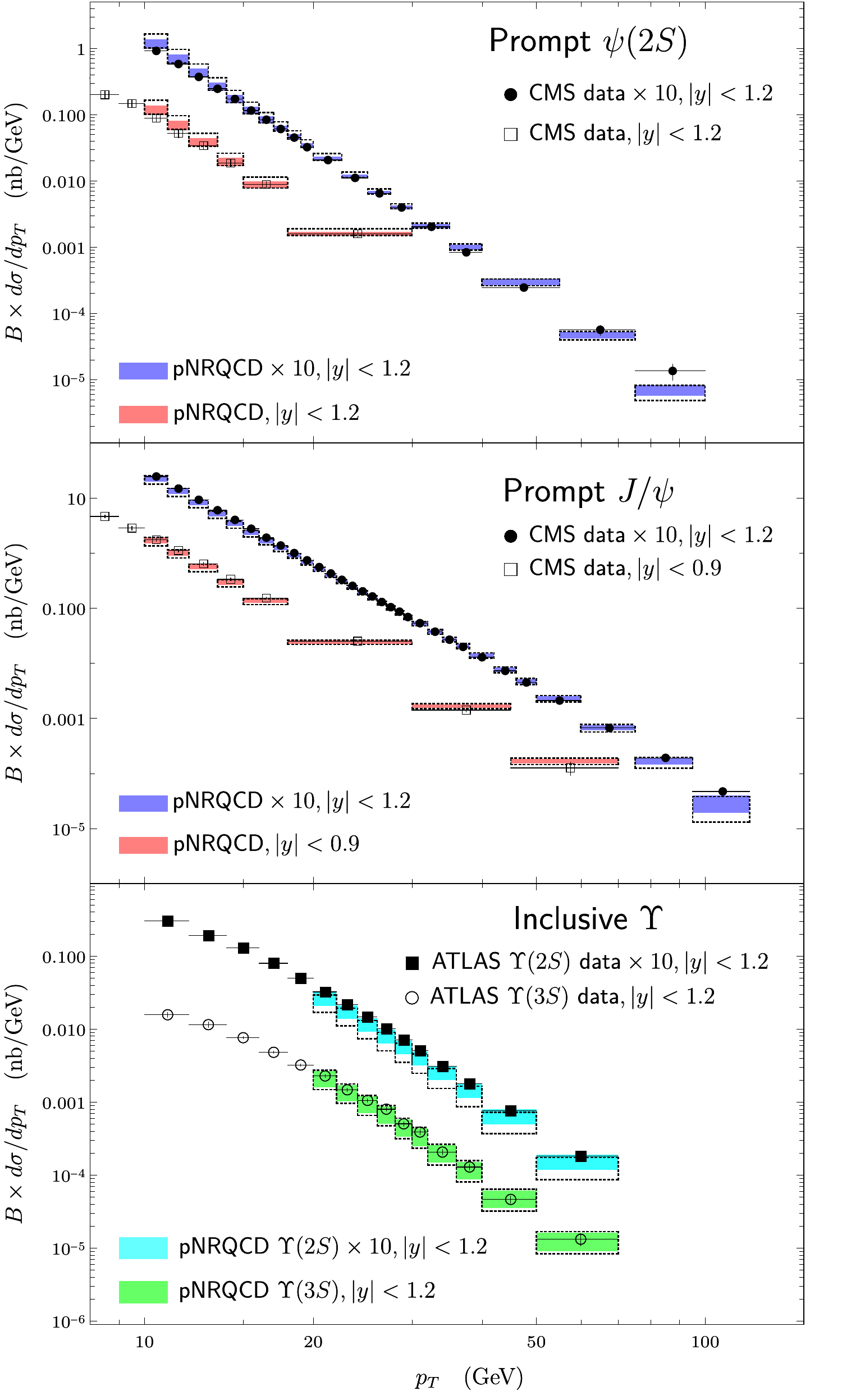}
\caption{\label{fig:rates}
The $p_T$-differential cross sections for $J/\psi$, $\psi(2S)$, 
$\Upsilon(2S)$, and
$\Upsilon(3S)$ at the LHC center of mass energy $\sqrt{s}=7$~TeV compared with the CMS and 
ATLAS measurements~\cite{CMS:2011rxs, ATLAS:2012lmu, CMS:2015lbl}.
Here, $B$ is the dilepton branching ratio.
For each quarkonium state, the dotted outlined bands are pNRQCD results 
obtained by excluding that quarkonium data from the fit.
}
\end{figure}

Our fits strongly constrain the value of ${\cal E}_{00}$ to be positive. 
This happens because, the ratios of the charmonium and bottomonium cross
sections at comparable values of $p_T/m$ depend mainly on the
ratios of $|R^{(0)}_V(0)|^2$, the quark mass, and the running of the
correlators. 
Since the running of ${\cal E}_{10;10}$ depends on ${\cal E}_{00}$,
the pNRQCD analysis determines rather precisely ${\cal E}_{00}$, 
and eventually the CO LDME $\langle {\cal O}^V(^3P_0^{[8]})\rangle$
as well.
This distinguishes the pNRQCD analysis from other existing approaches.
A positive ${\cal E}_{00}$ implies that the value of ${\cal E}_{10;10}$
at the scale $\Lambda = m$ is larger for bottomonium than for charmonium.
We expect that the sign of ${\cal E}_{00}$ will not change because of 
corrections of higher orders in $\alpha_s$, as radiative 
corrections shall affect the charmonium and bottomonium SDCs in a similar way.

In Fig.~\ref{fig:rates} we show our results for the prompt 
production rates of $J/\psi$ and $\psi(2S)$, and inclusive cross
sections of $\Upsilon(2S)$ and $\Upsilon(3S)$ at
the LHC center of mass energy $\sqrt{s}=7$~TeV 
compared
with CMS and ATLAS measurements from Refs.~\cite{CMS:2011rxs, ATLAS:2012lmu,
CMS:2015lbl}, which are used in our fits. 
The theoretical uncertainties are determined so that they
encompass the uncertainties in the correlators in the two $p_T$ regions. 
The pNRQCD results for the cross sections are in fair agreement with
measurements at large $p_T$. 
In Fig.~\ref{fig:rates} we also show, as dotted outlined bands, the 
pNRQCD results where the production rates for each quarkonium state are
obtained by excluding that quarkonium data from the fit. In all cases, the
results are consistent with the fits from all available data.

We note that the bulk of the cross section 
comes from the remnant of the cancellation between the $^3P_J^{[8]}$ and
$^3S_1^{[8]}$ channels ($\hat{\sigma}_{^3P_J^{[8]}}$ is negative at large
$p_T$, while $\hat{\sigma}_{^3S_1^{[8]}}$ is positive);
moreover, the contribution from the $^1S_0^{[8]}$ channel is small. 
We have tested the stability of our results against the large
cancellations between channels by imposing an upper $p_T$ cut, 
and found that it has negligible effects to our fits.

\begin{figure}[t]
\includegraphics[width=0.49\textwidth]{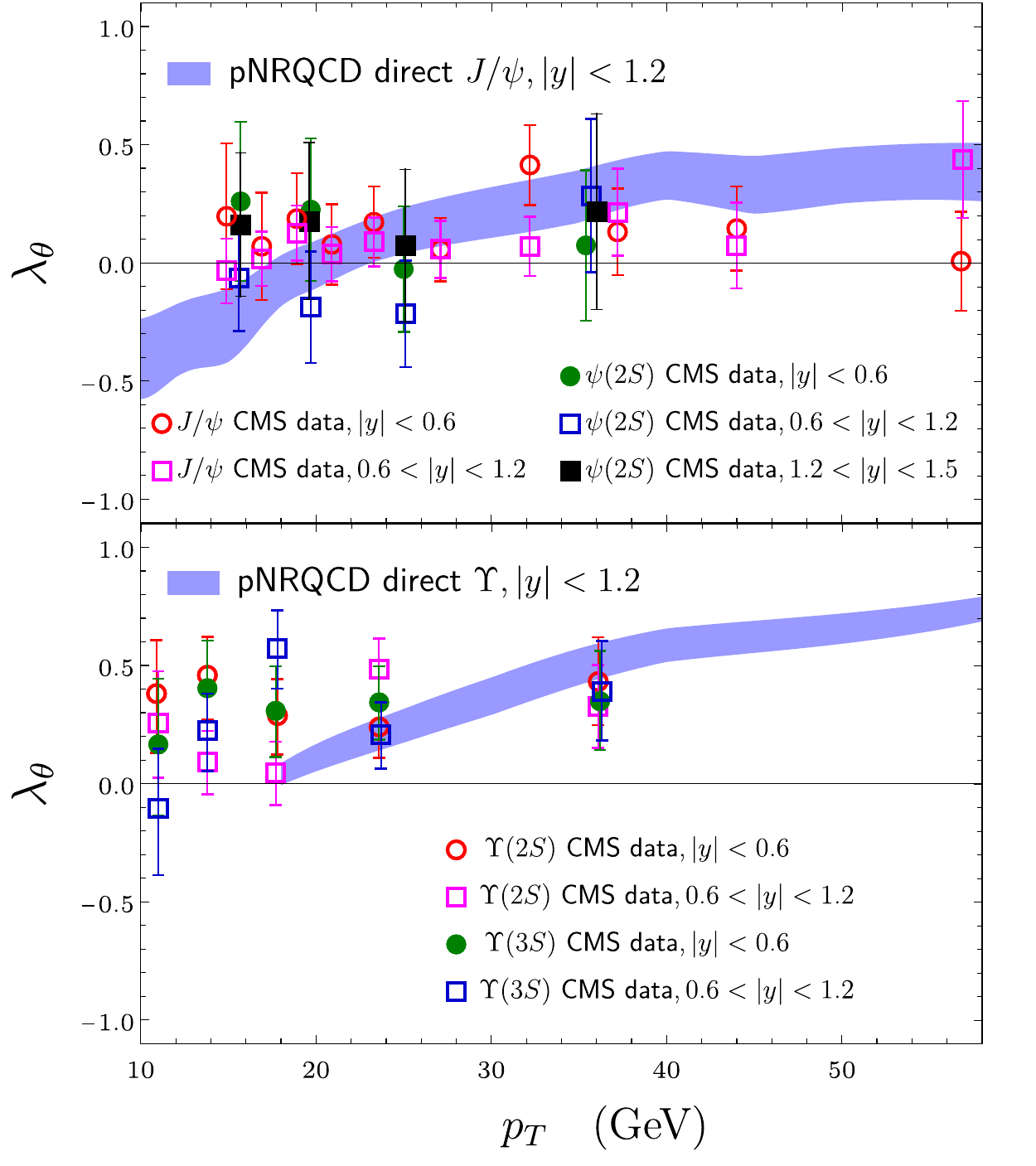}
\caption{\label{fig:pols}
The polarization parameter $\lambda_\theta$ in the helicity frame for direct
$J/\psi$, $\psi(2S)$, and $\Upsilon$ compared to CMS
measurements~\cite{CMS:2012bpf, CMS:2013gbz}. 
}
\end{figure}

An important observable that lets us put the CO
contributions to the test is the polarization of the 
quarkonium at large $p_T$. 
We consider the polarization parameter $\lambda_\theta$
in the helicity frame, which takes values $+1$, 0, and $-1$ when the quarkonium
is transverse, unpolarized, and longitudinal,
respectively~\cite{Beneke:1996yw, Beneke:1998re, Faccioli:2010kd}. 
The pNRQCD expressions for the CO LDMEs in
Eqs.~(\ref{eq:octet_result}) imply that the polarization of directly produced
quarkonium is independent of the radial excitation, up to corrections of
higher orders in $v^2$. Although the $^3S_1^{[8]}$ and $^3P_J^{[8]}$ channels
are strongly transverse, the large cancellation between the two channels
result in smaller values of $\lambda_\theta$. 
Because ${\cal E}_{00}$ is positive, 
we expect that $\Upsilon$ is more transverse than $J/\psi$ or $\psi(2S)$ 
at comparable values of $p_T/m$, due to the running of ${\cal E}_{10;10}$ 
which makes it take a greater value at the scale of the 
bottom quark mass compared to the charmonium case. 
These expectations are supported by the pNRQCD
results for $\lambda_\theta$ in the helicity frame shown in
Fig.~\ref{fig:pols}, which also agree with CMS
measurements~\cite{CMS:2012bpf, CMS:2013gbz}.

Other observables that provide tests of the CO LDMEs include production rates 
of $J/\psi$ at $ep$ and lepton colliders, and production of $\eta_c$. 
As have been pointed out in Refs.~\cite{Butenschoen:2012qr,
Butenschoen:2014dra}, 
many LDME determinations based on hadroproduction data are known to 
overestimate these cross sections
compared to the measurements in Refs.~\cite{LHCb:2014oii,
H1:2002voc, Belle:2009bxr, H1:2010udv}.
A small or negative value of $\langle {\cal O}^{J/\psi} (^1S_0^{[8]}) \rangle$,
similar to what we obtain from our result for $c_F^2 {\cal B}_{00}$,  
can reduce the size of these cross sections compared to 
existing hadroproduction-based approaches in Refs.~\cite{Chao:2012iv,
Gong:2012ug, Bodwin:2014gia, Bodwin:2015yma}, diminishing in this way
the tension with measurements~\cite{Brambilla:2022long}. 

The pNRQCD calculation of the NRQCD LDMEs of $S$-wave heavy quarkonia that we
have presented in this paper provides expressions for the color-singlet and
color-octet LDMEs in terms of quarkonium wave functions and universal gluonic
correlators. This brings in a reduction of the number of nonperturbative
unknowns and significantly enhances the predictive power of the 
factorization formalism for inclusive production of heavy quarkonia. 
The universality of the gluonic correlators lets us determine the LDMEs 
from charmonium and bottomonium data, which leads to strong constraints on
the LDMEs. Especially, the $P$-wave CO LDMEs are strongly constrained, 
which may be more robust against radiative corrections compared to existing
determinations. 
The pNRQCD results for the inclusive cross sections of $J/\psi$, $\psi(2S)$, 
and excited $\Upsilon$ states and their polarizations at the LHC are shown 
in Figs.~\ref{fig:rates} and \ref{fig:pols}. They 
agree with measurements at the LHC. 
More measurements of production rates of excited $\Upsilon$ states 
at large $p_T$ will 
help further reduce theoretical uncertainties. 
The pNRQCD calculation of the NRQCD LDMEs for production of $S$-wave heavy
quarkonia presented in this paper may be important in resolving
the longstanding puzzle of the heavy quarkonium production mechanism.

\medskip

\begin{acknowledgments}
We thank Peter Vander Griend for his careful reading of the manuscript.
The work of N.~B. is supported by the DFG (Deutsche Forschungsgemeinschaft,
German Research Foundation) Grant No. BR 4058/2-2. N.~B., H.~S.~C., and A.~V.
acknowledge support from the DFG cluster of excellence ``ORIGINS'' under
Germany's Excellence Strategy - EXC-2094 - 390783311. 
The work of A.~V. and X.-P.~W. is funded by the DFG Project-ID 196253076 - TRR 110.

\end{acknowledgments}

\bibliography{hadropro-Swave.bib}

\end{document}